\documentclass[%showpacs, 
superscriptaddress,twocolumn, preprintnumbers, nofootinbib,amsmath,amssymb]{revtex4}
\usepackage{graphicx}% Include figure files
\usepackage{dcolumn}% Align table columns on decimal point
\usepackage{bm}% bold math
\usepackage{xcolor}

\newcommand\ee{\end{equation}}
\newcommand\be{\begin{equation}}
\newcommand\eea{\end{eqnarray}}
\newcommand\bea{\begin{eqnarray}}

\newcommand\mpl{M_{\rm pl}}

\definecolor{DarkBlue}{rgb}{0.15,0.15,0.85}
\usepackage[linktocpage=true]{hyperref}
\hypersetup{
colorlinks=true,
citecolor=DarkBlue,
linkcolor=DarkBlue,
urlcolor=DarkBlue,
}

\usepackage{soul}
\newcommand{\Comment}[1]{{}}

\begin{document}

%\preprint{}

\title{$\phi^2$ or Not $\phi^2$: Testing the Simplest Inflationary Potential}% Force line breaks with \\

\author{Paolo Creminelli}
\affiliation{Abdus Salam International Centre for Theoretical Physics,\\ Strada Costiera 11, 34151, Trieste, Italy}
\affiliation{Institute for Advanced Study, Princeton, New Jersey 08540, USA}
\author{Diana L\'opez Nacir}
\affiliation{Abdus Salam International Centre for Theoretical Physics,\\ Strada Costiera 11, 34151, Trieste, Italy}
\affiliation{Departamento de F\'isica and IFIBA, FCEyN UBA, Facultad de Ciencias Exactas y Naturales, Ciudad Universitaria, Pabell\'on I, 1428 Buenos Aires, Argentina}
\author{Marko Simonovi\'c}
\affiliation{SISSA, via Bonomea 265, 34136, Trieste, Italy}
\affiliation{Istituto Nazionale di Fisica Nucleare, Sezione di Trieste, I-34136, Trieste, Italy}
\author{Gabriele Trevisan}
\affiliation{SISSA, via Bonomea 265, 34136, Trieste, Italy}
\affiliation{Istituto Nazionale di Fisica Nucleare, Sezione di Trieste, I-34136, Trieste, Italy}
\author{Matias Zaldarriaga}
\affiliation{Institute for Advanced Study, Princeton, NJ 08540, USA}


\begin{abstract}
The simplest inflationary model $V=\frac12 m^2\phi^2$ represents the benchmark for future constraints. For a quadratic potential, the quantity $(n_s-1)+r/4+11 (n_s-1)^2/24$ vanishes (up to corrections which are cubic in slow roll) and can be used to parametrize small deviations from the minimal scenario. Future constraints on this quantity 
will be able to distinguish a quadratic potential from a pseudo-Nambu-Goldstone boson with $f \lesssim 30 \mpl$ and set limits on the deviation from unity of the speed of sound $| c_s-1| \lesssim 3\times 10^{-2}$  (corresponding to an energy scale $\Lambda\gtrsim 2\times 10^{16}\, \mathrm{GeV}$), and on the contribution of a second field to perturbations ($\lesssim 6 \times 10^{-2}$). The limiting factor for these bounds will be the uncertainty on the spectral index. The error on the number of e-folds will be $\Delta N \simeq 0.4$, corresponding to an error on the reheating temperature $\Delta T_\mathrm{rh}/T_\mathrm{rh}\simeq 1.2$. We comment on the relevance of non-Gaussianity after BICEP2 results.
\end{abstract}

%\pacs{\dots}% PACS, the Physics and Astronomy
\maketitle

%%%%%%%%%%%%%%%%%%%%%%%%%%%%%%%%%%%%%%%%%%%%%%%%%%%%%%%%%
%%%%%%%%%%%%%%%%%%%%%%%%%%%%%%%%%%%%%%%%%%%%%%%%%%%%%%%%%
{\em Motivations.}---The recent detection of B-modes in the polarization of the cosmic microwave background (CMB) by BICEP2 \cite{Ade:2014xna} indicates a high level of primordial tensor modes. This requires \cite{Lyth:1996im} a large excursion of the inflaton during inflation $\Delta\phi \gtrsim \mpl$, which challenges the naive expectation that higher-dimension operators suppressed by powers of $\mpl$ spoil the slow-roll conditions. While, before BICEP2, the crucial question for inflation was ``large or small $r$?'' we are now facing a new dichotomy: ``$\phi^2$ or not $\phi^2$?" The two possibilities are qualitatively different. A large field model that is not quadratic, say $V \propto \phi^{2/3}$, suggests an interesting UV mechanism, such as monodromy inflation \cite{Silverstein:2008sg}, for instance. If data will, on the other hand, favor a quadratic potential, the simplest explanation will be that inflation occurs at a generic minimum of a potential whose typical scale of variation $f$ is much larger than the Planck scale. Indeed, an approximate shift symmetry gives rise to potentials that are periodic in $\phi/f$, such as, for instance, $V = \Lambda^4 (1-\cos(\phi/f))$ \cite{Freese:1990rb,Freese:2014nla}. For $f \gg \mpl$, inflation occurs near a minimum of the potential, where one can approximate $V\propto\phi^2$. In string theory, it seems difficult to obtain a parametric separation between $f$ and $\mpl$, although there is no issue at the level of field theory \cite{ArkaniHamed:2003wu,Kaloper:2011jz}. Therefore, if quadratic inflation will remain compatible with the data, it will be important to study small deviations from it, to understand to which extent the quadratic approximation holds and to limit other possible deviations from the simplest scenario of inflation.

\begin{figure}[!!!h]
\begin{center}
\includegraphics[width=0.485 \textwidth]{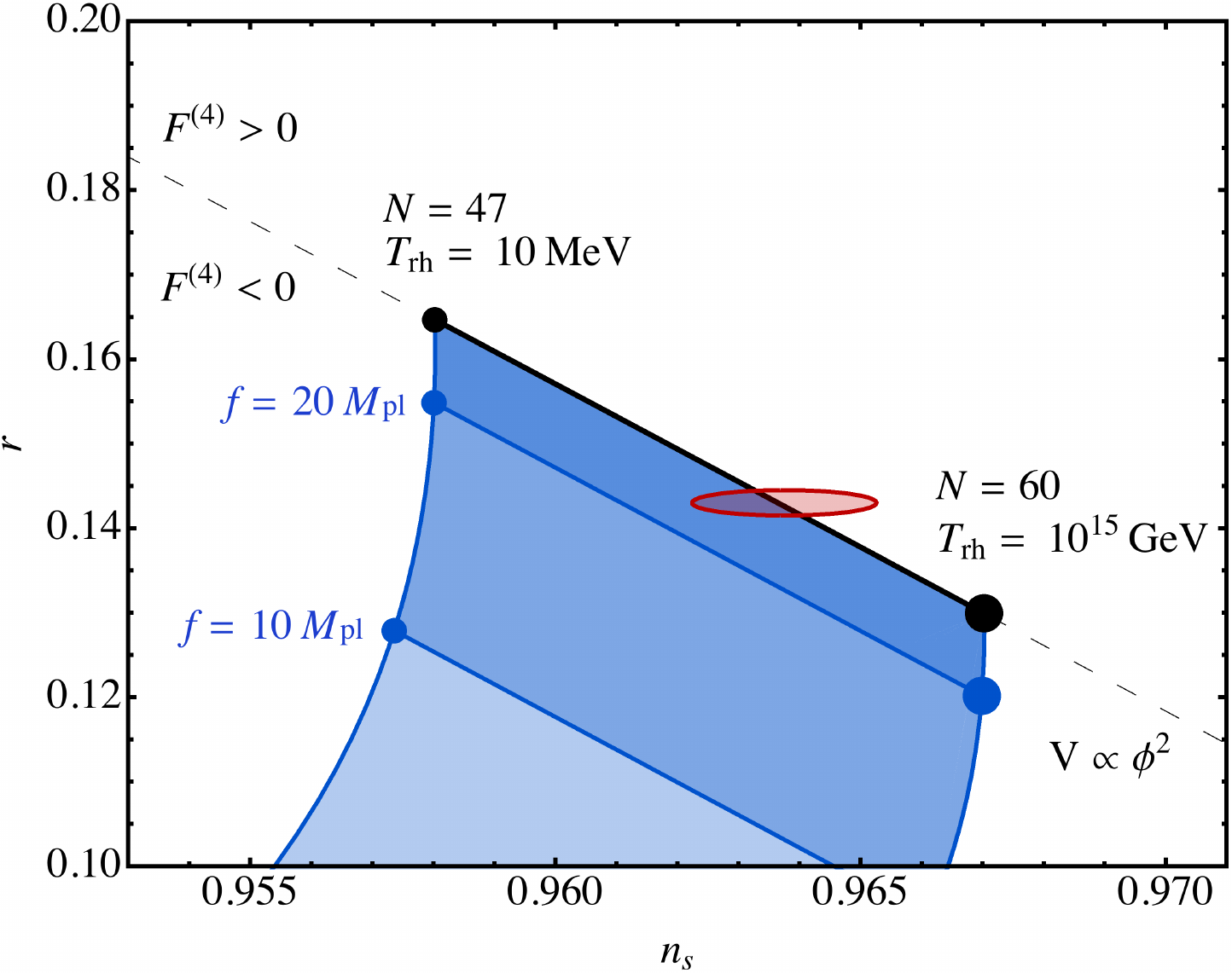}
\end{center}
\caption{\small {\em Future constraints on $f$ assuming a simple cosine potential. The dashed curve corresponds to Eq.~\eqref{eq.GoldenRule}, and the black segments cover the interval of reheating temperatures $T_{\mathrm{rh}} \in [10\mathrm{\;MeV},10^{15}\mathrm{\;GeV}]$. A wider range of $N$ is allowed if one considers nonstandard cosmological evolutions after inflation. Red $1\sigma$ contour corresponds to a futuristic measurement with $\sigma_{n_s-1}=\sigma_r=10^{-3}$, compatible with a  quadratic potential. All quantities are evaluated at $k=0.002 \; \mathrm{Mpc}^{-1}$.}}
\label{fig:ns-r-plot}
\end{figure}

Inflationary predictions must face our ignorance about the reheating process and the subsequent evolution of the Universe. All this is encoded in the number of e-folds $N$ between when the relevant modes exit the horizon and the end of inflation. The dependence on $N$ is rather strong (see Fig.~\ref{fig:ns-r-plot}) and it will become larger than the experimental sensitivity on $n_s$ and $r$.  To study small deviations from $V = \frac12 m^2\phi^2$, we have to concentrate on a combination of observables that does not depend on $N$ (\footnote{The power spectrum normalization fixes $m\simeq 1.5\times10^{13}\,\mathrm{ GeV}$, $V\simeq (2\times10^{16}\,\mathrm{GeV})^4$ and $\Delta\phi\simeq 15\,\mpl$, assuming $N=60$.}). At linear order in $1/N$, given that for a quadratic potential $n_s-1 = -2/N$ and $r = 8/N$, a prediction that is independent of $N$ is obviously $(n_s-1)+r/4=0$. Since corrections at second order in slow roll will not be completely negligible in the future, it is worthwhile to go to order $1/N^2$. With the use of the explicit formulas at second order in slow roll \cite{Stewart:1993bc}, it is straightforward to verify\footnote{Up to second order in slow roll we have 
%\be
%n_s - 1= - 4\epsilon -\left(8C+6\right) \epsilon^2, \quad r=4 \epsilon  \left(1-\frac{2 \epsilon}{3}+2 \epsilon C  \right)\,, \nonumber
%\ee
\begin{align}
n_s-1 &=2 \eta -6 \epsilon - 2C(12\epsilon ^2 + \xi ) + \frac{2}{3}(\eta^2-5\epsilon^2 +\xi)+(16 C-2) \eta \epsilon \;,\nonumber\\
r &= 16 \epsilon  \left[1-\frac{4 \epsilon}{3}+\frac{2 \eta}{3}+2 C (2\epsilon-\eta)  \right]\,,\nonumber
\end{align}
where  $C\equiv-2+\ln 2 + \gamma$, with $\gamma=0.57721\ldots$ the Euler-Mascheroni constant, and the slow-roll parameters are defined as
\begin{equation}
\epsilon\equiv\frac{\mpl^2}{2} \left(\frac{V'}{V}\right)^2\, , \ \ \ \eta\equiv\mpl^2\frac{V''}{V}\,, \ \ \ \xi\equiv\mpl^4 \frac{V'''V'}{V^2}\,. \nonumber
\end{equation}
} 
that for a quadratic potential
\begin{equation}\label{eq.GoldenRule}
\begin{split}
(n_s &-1) +\frac{r}{4} + \frac{11}{24}(n_s-1)^2 =0\,,
\end{split}
\end{equation}
up to corrections of order $N^{-3}$ (\footnote{Up to $1/N^3$ corrections we can equivalently write $(n_s-1)+r/4+11/384 \cdot r^2 = 0$. This form can be useful in future given that the error on $(r/4)^2$ is expected to be smaller than the one on $(n_s-1)^2$.}), which we can safely ignore. Assuming that data will favour a $\phi^2$ potential we can use the equation above to study how sensitive we will be to small departures from the simplest  scenario. If we take the measurement of the tilt from Planck \cite{Ade:2013ydc}  $n_s-1=-0.0397\pm 0.0073$ and the recent value of $r$ measured by BICEP2 \cite{Ade:2014xna} $r=0.20_{-0.05}^{+0.07}$, the lhs of Eq.~(\ref{eq.GoldenRule}) is equal to $0.01 \pm 0.02$ 
$(\footnote{This number has to be taken with great caution because it does not include foreground subtraction in BICEP2 result. If one uses Planck data only, the lhs of Eq.~\eqref{eq.GoldenRule} is compatible with zero at $\sim 2\sigma$. })$.
Optimistically we can assume we will be able to measure $r$ with a precision of $1\%$ \cite{Dodelson:2014exa}. Regarding $n_s-1$, future experiments such as EUCLID \cite{Amendola:2012ys} or PRISM \cite{Andre:2013afa} should be able to go down to a $10^{-3}$ error. Therefore, the uncertainty on the quantity above will be $\sim 10^{-3}$, dominated by the error on the spectral index. Notice that different experiments are sensitive to different scales $k$. Given that Eq.~\eqref{eq.GoldenRule} is independent of $N$, it is valid on any scale provided that both $n_s$ and $r$ are evaluated at the same $k$. Therefore, it is important to keep in mind that the experimental results have to be properly combined at the same scale\footnote{This is the case also for BICEP2 and Planck, but with the current errors this difference is negligible.}. 

Let us now study what these futuristic limits will imply for deviations from the simplest model of the Universe.

\vspace{.3cm}
%%%%%%%%%%%%%%%
{\em Pseudo-Nambu-Goldstone boson potential.}---A PNGB has a potential of the form $V=\Lambda^4 F(\phi/f)$ (\footnote{In the extra-dimensional model of Ref.~\cite{ArkaniHamed:2003wu}, the explicit form of $F$ depends on the number of particles, their charges, masses, and boundary conditions.}), where $\Lambda$ is the scale of breaking of the approximate shift symmetry, $F$ is a periodic function and $f$ is the decay constant. 
The simplest example is given by
\be
\label{natural}
V(\phi) = \Lambda^4 \left[ 1 - \cos \left(\frac \phi f \right) \right]\;,
\ee
where $f$ has to be bigger than $\mpl$ in order for the slow-roll conditions to be satisfied and for very large $f \gg \mpl$ the model becomes indistinguishable from a $\phi^2$ potential. For this potential, Eq.~\eqref{eq.GoldenRule} will not be exactly zero. It is easy to calculate the leading correction in slow roll and for $\mpl/f \ll 1$
\be
\label{ns-r-natural}
n_s -1 = -\frac{2}{N} + {\cal O}\left(\frac{\mpl}{f}\right)^4\,, \ \ \ \ \ r = \frac{8}{N} -4 \left(\frac{\mpl}{f}\right)^2\;.
\ee
This gives a correction to Eq.~\eqref{eq.GoldenRule}
\be
(n_s-1) +\frac{r}{4} + \frac{11}{24}(n_s-1)^2= -   \left(\frac{\mpl}{f}\right)^2 \;.
\ee
If the error on the lhs is of order $10^{-3}$, this translates into the limit $f \gtrsim 30 \mpl$. This would convincingly suggest there is a parametric separation between the two scales, which the UV theory would have to address.
To illustrate this point, in Fig.~\ref{fig:ns-r-plot} we present a plausible forecast for the future observations in the $(n_s,r)$ plane together with the predictions of natural inflation for different values of $f$.

For a generic $F$ expanding around the minimum we get
\be
V(\phi) = \Lambda^4 \left( \frac{1}{2} \frac{\phi^2}{f^2} + \frac{F^{(3)}}{6} \frac{\phi^3}{f^3} + \frac{F^{(4)}}{24} \frac{\phi^4}{f^4} +\cdots \right)\;.
\ee
For the following analysis, we assume $F^{(n)}$ to be of order 1. For the moment, let us assume the function $F$ is symmetric around the minimum. Notice that with positive $F^{(4)}$ we can get $n_s$ and $r$ above the $m^2\phi^2$ curve, unlike in the case of a simple cosine potential (see Fig.~\ref{fig:ns-r-plot}). At leading order in slow roll
\be
(n_s -1) +\frac r 4 + \frac{11}{24} (n_s -1) ^2 = F^{(4)} \left( \frac{\mpl}{f}\right)^2\:,
\ee
and one can constrain the combination on the rhs $f/\sqrt{|F^{(4)}|}\gtrsim 30 \mpl$. Therefore, for $F^{(4)}$ of order one, this does not change the lower bound on $f$ significantly. 

If we now allow for nonzero $F^{(3)}$ and the cubic term dominates, Eq.~\eqref{eq.GoldenRule} reads
\be
(n_s -1) +\frac r 4 + \frac{11}{24} (n_s -1) ^2 = \pm\frac 23 \sqrt{2\epsilon}\;F^{(3)} \frac{\mpl}{f} \;,
\ee
where the sign depends on whether inflation occurs for positive or negative values of $\phi$.
The constraint on the rhs imposes\footnote{Here and in the following estimates, to be conservative, we use the minimal value of $\epsilon$ that corresponds to the maximal number of e-folds.} $f/F^{(3)}\gtrsim 86 \mpl$. Notice that in this case the lower bound on $f$ is even stronger.

\vspace{.3cm}
%%%%%%%%%%%%%%%
{\em General deviations from $\phi^2$.}---One can use the same technique to constrain other deviations from the simplest scenario: they will all contribute to the rhs of Eq.~\eqref{eq.GoldenRule}. Let us first focus on small deviations from $m^2\phi^2$ coming from the shape of the potential (see for example, Refs. \cite{Kallosh:2014xwa, Harigaya:2014qza}). It is straightforward to obtain the corrections to Eq.~\eqref{eq.GoldenRule} for a generic $V(\phi)$ up to second order in slow-roll parameters
\begin{align}
\label{general}
(n_s& -1) +\frac r 4 + \frac{11}{24} (n_s -1) ^2 = -2(\epsilon-\eta) \;.
\end{align}
Notice that on the rhs of Eq.~(\ref{general}) we keep only the first nonvanishing correction. 

Another kind of corrections comes from derivative interactions. Indeed, from the effective field theory point of view quantum corrections will generate higher-dimensional operators suppressed by some scale $\Lambda$. Particularly important are the operators compatible with an approximate shift symmetry for $\phi$. For example, a term of the form $(\partial \phi)^4/\Lambda^4$ in the Lagrangian corresponds to a correction to the speed of propagation of the perturbations 
\be
c_s^2 -1= 16 \frac{\dot H \mpl^2}{\Lambda^4}\;.
\ee
Therefore, constraints on the speed of sound transfer into constraints on $\Lambda$.
In models with $c_s<1$, it is important to stress that $r=16\epsilon c_s$, whereas $n_s-1$ is independent of $c_s$ (it only depends on it through $s\equiv\dot c_s/H c_s$). In the absence of cancellations, the current value for $n_s-1$ and the detection of a high level of primordial tensor modes imply that $c_s$ cannot be much smaller than $1$.

For the case of a quadratic potential, one can quantify the bounds on $c_s$ more precisely in a way that is insensitive to $N$. The correction to Eq.~(\ref{eq.GoldenRule}) reads
\begin{align}
\label{cs}
(n_s &-1) +\frac r 4 + \frac{11}{24} (n_s -1) ^2 =-s+ \frac r 4 \left(1-\frac{1}{c_s}\right)\;.
\end{align}
If the total error on the lhs is of the order $10^{-3}$,  $|c_s-1|$ is constrained to be $\lesssim 3 \times 10^{-2}$. In particular, we can put a lower bound on the energy scale $\Lambda$ to be $\Lambda \gtrsim 2\times 10^{16} \;\mathrm{GeV}$ which is as high as the inflationary scale. 

Another way to constrain $c_s$ is to use the standard consistency relation for the tilt of tensor modes $n_T$ 
\be\label{eq.cr2}
r+8n_T = \left( 1- \frac{1}{c_s} \right)r \;.
\ee
This relation has the major advantage of being valid for any potential, but it is difficult to imagine we will be able to verify it with significant precision. Given that from CMB experiments it will be hard to measure $n_T$ with a precision better than $\Delta n_T \sim 0.1$, the constraint on $c_s$ is weaker than the one obtained above. However, in the very far future we might be able to constrain $r$ and $n_T$ much better by the detection of primordial gravitational waves with interferometers \cite{Caligiuri:2014sla}. Optimistically, the error on $n_T$ could be as low as $5 \times^{-3}$ and the relation of Eq.~(\ref{eq.cr2}) could constrain $c_s$ even better than Eq.~(\ref{cs}).

Another possible departure from the simplest model is the presence of a subdominant component in the spectrum due to a second field. In these models (curvaton, modulated reheating, etc.) inflation is driven by the inflaton, but a second scalar field $\sigma$ is contributing to the curvature perturbation with a fraction
\begin{equation}
\begin{split}
q\equiv\frac{P^\sigma_\zeta}{P^\phi_\zeta+P^\sigma_\zeta}\;,
\end{split}
\end{equation}
where $P^x_\zeta$ is the contribution of the field $x$ to the power spectrum of the curvature perturbation $\zeta$. The correction to Eq.~(\ref{eq.GoldenRule}) up to first order in slow roll is 
\begin{align}
\label{curvaton}
(n_s -1) +\frac r 4 + \frac{11}{24} (n_s -1) ^2= q \left( -\frac{r}{8} + \frac 23 \frac {V_\sigma''}{H^2} \right) \;.
\end{align}
%Notice that corrections proportional to $q \,\epsilon$ exactly cancel. 
Assuming that the error on the lhs of Eq.~(\ref{curvaton}) is $10^{-3}$, this relation constraints $q\lesssim 0.06$.

One may consider the case in which different corrections to the rhs of Eq.~\eqref{eq.GoldenRule} cancel, so that we accidentally get the same predictions as the $\phi^2$ model. In this case, one can hope to break the degeneracy by looking at the running of the power spectrum. For a quadratic potential $\alpha = -(n_s-1)^2/2=-r^2/32\simeq 8\times10^{-4}$. 

\vspace{.3cm}
%%%%%%%%%%%%%%%
{\em Constraints on $N$.}---So far, we have focused on a combination of observables that is $N$ independent. On the other hand, for a quadratic potential one will also get a good constraint on the number of e-folds. With the numbers quoted above, the best constraint will most likely come from $r$, which will give $\Delta N  \simeq 0.4$. This translates into an error on the reheating temperature
\begin{equation}
\begin{split}
\frac{\Delta T_\mathrm{rh}}{T_\mathrm{rh}}\simeq 1.2 \;,
\end{split}
\end{equation}
assuming we know the evolution after reheating. Notice that while it is easy to reduce $N$ (longer reheating, periods of matter domination or phase transitions in the late universe, large number of relativistic degrees of freedom $g_*$), the upper bound on $N$ corresponding to instantaneous reheating and conventional cosmological evolution is very robust. In some sense, it corresponds to the very simplest Universe. 

\vspace{.3cm}
%%%%%%%%%%%%%%%
{\em What if not $\phi^2$?}---All the discussion so far concentrated on $\phi^2$ inflation. If nature has chosen another monomial potential $V \propto \phi^p$, we can still build an observable which does not depend on $N$. It is easy to get
\be
\label{eq:p}
(n_s-1)+\frac{2+p}{8p}r +\frac{3p^2+18p-4}{6(p+2)^2}(n_s-1)^2 = 0 \;.
\ee 
As before we will have errors of order $10^{-3}$ on this expression\footnote{Notice that in eq.~\eqref{eq:p} as we go to lower values of $p$ the coefficient of $r$ increases: the tensor contribution becomes more and more important to discriminate the model.}. It is straightforward to generalize Eqs.~(\ref{cs}) and (\ref{curvaton}) to analyze the constraints on the speed of sound or the presence of a curvaton component. 
%For ``normal" powers, say $V \propto \phi$, we will be quite convinced we have found the correct model of inflation\footnote{Notice that in Eq.~\eqref{eq:p} as we go to lower values of $p$ the coefficient of $r$ increases: the tensor contribution becomes more and more important to discriminate the model.}. Of course if $p$ is allowed to vary continuously, one cannot reach any firm conclusion. However, from the model building point of view some (fractional) powers will be always preferred. We can quantify deviations from the given preferred power $p$ looking at deformations of the potential of the form $V\propto \phi^{p+\alpha}$. In this case the relation \eqref{eq:p} at linear order in $\alpha$ becomes
%\be
%(n_s-1)+\frac{2+p}{8p}r +\frac{3p^2+18p-4}{6(p+2)^2}(n_s-1)^2 = \frac{\alpha r}{4 p^2} \;.
%\ee
One can also invert Eq.~(\ref{eq:p}) to find the allowed range of $p$. This reads
\begin{equation}
\begin{split}
p=&-\frac{2 r}{8 (n_s-1)+r}\\
&-\frac{64 (n_s-1)^3}{(8(n_s-1)+r)^2}+(n_s-1)-\frac{7}{24}r\:.
\end{split}
\end{equation}
For ``normal" powers we will be quite convinced we have found the correct model of inflation. With the current errors  the bounds on $p$ are too loose to be interesting, but this may change in the future. For example, for a linear potential the error will be $\Delta p \simeq0.06$, and this will allow discrimination of this model from $\phi^{2/3}$.
%The current values of $r$ and $n_s$ give $p=3.6\pm3.2$, but for future measurements the error will shrink down by a factor of 10.
%The constraint on $\alpha$ depends on $p$. For $p=2/3$ we would get $|\alpha|\lesssim 0.01$ while for $p=3$ the constraint would be $|\alpha|\lesssim 0.2$.

\vspace{.3cm}
%%%%%%%%%%%%%%%
{\em Non-Gaussianity.}---So far our discussion has concentrated on the power spectra: what about higher-order correlation functions? As discussed above, in single-field models (independently of the potential) $r$ is suppressed by $c_s$, so that the speed of sound cannot be much smaller than 1. 
Therefore, the cubic operator related by symmetry to $c_s$ \cite{Cheung:2007st} cannot give sizeable non-Gaussianities since $f_{\mathrm{NL}}^\mathrm{eq}\sim1/c_s^2$  (the Planck constraint \cite{Ade:2013ydc} is $c_s\ge 0.02$). However the second independent operator $\dot\pi^3$ can still be large. It is straightforward to check that this situation is radiatively stable \cite{Baumann:2011nk}: loops induce order-one corrections to the speed of sound.
Moreover, the three-point function can be large for $c_s =1$ if it is generated by operators with more than one derivative per field \cite{Creminelli:2010qf}.  Another possibility is that the four-point function is large, while the bispectrum is suppressed: this can happen in a technically natural way as studied in Refs.~\cite{Senatore:2010jy}.  Non-Gaussianities are also relevant if scalar and tensor perturbations are both produced through particle creation involving dissipative effects \cite{LopezNacir:2011kk, Senatore:2011sp}.

It is interesting that, if we focus on a $\phi^2$ potential, the limits on $c_s$ discussed above can be far better than what is measurable through the three-point function.
%, even though in the $n_s-r$ plane the models could look like $m^2\phi^2$, one can discriminate them by looking at the non-Gaussianities.

Also, in multifield models $r$ is always suppressed compared to the single-field case (by a factor of $q$ assuming no mixing). It is, thus, unlikely that perturbations are dominated by a second field. However, when the perturbations due to the second field become very non-Gaussian, they induce a large observable non-Gaussianity $f_{\rm NL} \simeq 10^5 q^{3/2}$, even when they are subdominant in the power spectrum \cite{Senatore:2010wk}. Notice that the shape of non-Gaussianity can vary from local to equilateral if we consider general quasi-single-field models \cite{Chen:2009we, Craig:2014rta}. In conclusion, non-Gaussianities remain a powerful probe of inflation.

\vspace{.3cm}
%%%%%%%%%%%%%%%
{\em Conclusions.}---Any experimental result on the $(n_s, r)$ plane can be explained with a proper choice of the slow-roll parameters $\epsilon$ and $\eta$. On the other hand a particular curve on this plane stands out since it corresponds to the prediction of $V \propto \phi^2$, varying the number of e-folds $N$. Assuming data will remain compatible with this simple scenario, we studied the constraints we will be able to set on various deviations from the benchmark model.

\vspace{.3cm}
%%%%%%%%%%%%%%%
{\em Acknowledgements.}---It is a pleasure to thank M. Mirbabayi and in particular N.~Arkani-Hamed for useful discussions and comments on the draft. P.C. acknowledges the support of the IBM Einstein Fellowship. D.L.N., M.S., and G.T. are grateful to the IAS for hospitality during their work on this project. M.Z. is supported in part by  NSF Grants No. PHY-0855425, and No. AST-0907969, PHY-1213563 and by the David and Lucile Packard Foundation.

\end{document}